\documentclass[useAMS, usenatbib, usegraphicx]{mn2e}
\usepackage{times}
\usepackage{color}
\usepackage{amssymb}

\title[Spiral arm and inter-arm star formation]{Star formation in Galactic spiral arms and the inter-arm regions}
\author[D. J. Eden et al.]{D. J. Eden$^{1}$\thanks{E-mail: dje@astro.livjm.ac.uk}, T. J. T. Moore$^{1}$, L. K. Morgan$^{1}$, M. A. Thompson$^{2}$ and J.S. Urquhart$^{3}$ \\
$^{1}$Astrophysics Research Institute, Liverpool John Moores University, Twelve Quays House, Egerton Wharf, Birkenhead CH41 1LD\\
$^{2}$Centre for Astrophysics Research, Science \& Technology Research Institute, University of Hertfordshire, College Lane, Hatfield, Herts AL10 9AB\\
$^{3}$Max-Planck-Institut f\"{u}r Radioastronomie, Auf dem H\"{u}gel 69, 53121 Bonn, Germany}
\begin{document}

\date{Accepted. Received; in original form}

\pagerange{\pageref{firstpage}--\pageref{lastpage}} \pubyear{2012}

\maketitle

\label{firstpage}

\begin{abstract}

The line of sight through the Galactic Plane between longitudes $\emph{l}$ = 37.83$\degr$ and $\emph{l}$ = 42.50$\degr$ allows for the separation of Galactic Ring Survey molecular clouds into those that fall within the spiral arms and those located in the inter-arm regions. By matching these clouds in both position and velocity with dense clumps detected in the mm continuum by the Bolocam Galactic Plane Survey, we are able to look for changes in the clump formation efficiency (CFE), the ratio of clump-to-cloud mass, with Galactic environment. We find no evidence of any difference in the CFE between the inter-arm and spiral-arm regions along this line of sight. This is further evidence that, outside the Galactic Centre region, the large-scale structures of the Galaxy play little part in changing the dense, potentially star-forming structures within molecular clouds.

\end{abstract}


\section{Introduction}

The role played by spiral arms in triggering or regulating star formation is ambiguous. Significant increases in the efficiency of molecular-cloud formation from the neutral ISM in the spiral-arm entry shock has been observed \citep{Heyer98} and explained theoretically \citep{Dobbs06}. Since such shocks will be stronger at Galactocentric radii inside the co-rotation radius (which for the Milky Way is thought to be at around 8 kpc, \citealt{Lepine11}), it has been suggested that supernovae, rather than spiral structure, may be the dominant mechanism in regulating the state of the ISM and, hence, the mode, rate and efficiency of star formation in the outer Galaxy \citep{Dib09}.

Other theoretical predictions suggest that spiral arms may be largely organising features which mainly slows down the ISM gas in its orbit, but that this may allow larger giant molecular clouds (GMCs) to form \citep{Dobbs11}. A study by \citet{Roman-Duval10} found similar results using data from the Galactic Ring Survey (GRS), implying that clouds in inter-arm regions dissipate more quickly. If true, this may affect the mass function of stellar clusters that form, since radiative heating should suppress fragmentation in higher-column-density clouds without significantly affecting the overall star-formation rate or efficiency \citep{Krumholz10}. \citet{Moore12} found that around 70 per cent of the increase in star-formation rate density in spiral arms in the Galaxy is due to simple crowding. The remainder can either be due to rises in star-formation efficiency (SFE) or increases in the mean luminosity of massive YSOs.

These models and results are apparently contradicted, however, by other evidence showing spiral arms have little affect. For instance, \citet{Foyle10} found little difference in either the ratio of molecular gas to H$\,\textsc{I}$ or the star-formation efficiency in and out of the arms of two external spiral galaxies. \citet{Eden12}, examining the fraction of molecular gas in dense clumps within clouds,  found no difference between clouds in the Scutum-arm tangent and those in the foreground and background structures on the same line of sight.  This implies a constant conversion of molecular gas into dense, star-forming structures regardless of environment. 

To determine the effect of Galactic spiral arms on star formation, we need a model of the Galactic structure. The consensus, from the mapping of distances to observed H$\,\textsc{II}$ regions, is that the spiral structure of the Galaxy can be represented by a four-arm model (e.g. \citealt{Russeil03}; \citealt*{Paladini04}) but the geometry of these arms is not agreed upon. The four main arms - Norma, Sagittarius, Perseus and Scutum--Centaurus - are added to by the Near and Far 3-kpc arms. The Milky Way also has a central bar which can be split into a 3.1 - 3.5 kpc Galactic Bar at an angle of 20$\degr$ with respect to the Galactic Centre-Sun axis \citep{Binney91, Blitz91, Dwek95} and a non-symmetric structure, the ÔLong BarÕ \citep{Hammersley00}, at an angle of 44$\degr$ $\pm$ 10$\degr$ with a Galactic radius of 4.4 $\pm$ 0.5 kpc, as revealed by star counts from the Spitzer GLIMPSE survey \citep{Benjamin05}.

In this paper, we compare the fraction of molecular gas mass in dense, potentially star-forming clumps in Galactic spiral arms to that of clouds in the inter-arm zones. This is the clump-formation efficiency (CFE),  which is the dense-clump analogue (or precursor) of the SFE. The region covered by this study is the slice of the Galactic plane $\emph{l}$ = (37.83$\degr$-42.50$\degr$), $\mid$\emph{b}$\mid$ $\leq$ 0.5$\degr$, which will hereafter be referred to as the $\emph{l}$ = 40$\degr$ region. This line of sight is located between the Scutum--Centaurus tangent (at $\emph{l}$ $\approx$ 30$\degr$) and the Sagittarius tangent ($\emph{l}$ $\approx$ 50$\degr$), and intersects the Sagittarius arm twice. The $\emph{l}$ = 40$\degr$ region is suitable for comparisons between spiral-arm and inter-arm star formation, as multiple zones of each type are observed along the line of sight but there is no arm tangent, meaning that the arms are well separated. Populations of molecular clouds can be attributed to two intersections of the Sagittarius arm, the edge of the Scutum--Centaurus tangent and the Perseus arm, with corresponding inter-arm regions between. The inter-arm region between the Scutum--Centaurus and Sagittarius arms falls at the tangent point in this line of sight, so is not subject to distance ambiguities which can affect the results. This line of sight is also a molecular-rich inter-arm region \citep{Sawada12}. In the next section (Section 2), we give a brief overview of the data sets used. Methods for assigning distance to BGPS sources which cannot be associated with GRS clouds are described in Section 3. Section 4 contains the results and analysis and Section 5 is the discussion of the results. Section 6 is a summary of the conclusions. 

\section{Data Sets and Observations}

\subsection{Galactic Ring Survey}

The GRS \citep{Jackson06} mapped $^{13}$CO $J$ = 1 $\rightarrow$ 0 emission in Galactic longitude from $\emph{l}$ = 18$\degr$ to 55.7$\degr$ and $\mid$\emph{b}$\mid$ $\leq$ 1$\degr$, covering a total area of 75.4 deg$^{2}$, with a velocity range of --5 to 135 km s$^{-1}$ for $\emph{l}$ $\leq$ 40$^{\circ}$ and --5 to 85 km s$^{-1}$ for $\emph{l}$ $>$ 40$^{\circ}$ at an rms sensitivity of $\sim$ 0.13 K. The GRS is fully sampled with a 46$\arcsec$ angular resolution on a 22$\arcsec$ grid and has a spectral resolution of 0.21 km s$^{-1}$. The velocity range limits detections to within the Solar Circle.

A catalogue of 829 molecular clouds within the GRS region, identified using the CLUMPFIND algorithm \citep*{Williams94} was published by \citet{Rathborne09}. \citet{Roman-Duval09} determined distances to 750 of these clouds using H$\, \textsc{I}$ self-absorption (HISA) to resolve the kinematic distance ambiguities. This cloud distance catalogue was complemented by the work of \citet{Roman-Duval10} who made use of $^{12}$CO $J$ = 1 $\rightarrow$ 0 emission from the University of Massachusetts--Stony Brook survey \citep{Clemens86, Sanders86} to derive the masses, as well as other physical properties, of 580 molecular clouds. A power-law relation between their radii and masses was produced to allow the masses for a further 170 molecular clouds to be estimated. The associated cloud mass uncertainties are also catalogued. The clouds catalogued are those made up of the lower density, more diffuse molecular material within the ISM.

The molecular cloud mass completeness limit of the GRS as a function of distance is $M_{min}$ = $50 d^{2}$ M$_{\odot}$, where $d$ is the distance in kpc. Therefore the survey is complete above a mass of 4 $\times$ 10$^{4}$ M$_{\odot}$ out distances of 15 kpc \citep{Roman-Duval10}, so is believed to be complete for the distances probed by this study.

\subsection{The Bolocam Galactic Plane Survey}

The Bolocam Galactic Plane Survey (BGPS; \citealt{Aguirre11}) mapped 133 deg$^{2}$ of the north Galactic plane in the continuum at 271.1\,GHz ($\lambda = 1.1$\,mm) with a bandwidth of 46\,GHz, an rms noise level of 11--53\,mJy beam$^{-1}$ and an effective angular resolution of 33$\arcsec$. The survey was continuous from $\emph{l}$ = -10.5$\degr$ to 90.5$\degr$, $\mid$\emph{b}$\mid$ $\leq$ 0.5$\degr$ with cross-cuts which flare out to $\mid$\emph{b}$\mid$ $\leq$ 1.5$\degr$ at $\emph{l}$ = 3$\degr$, 15$\degr$, 30$\degr$ and 31$\degr$ and towards the Cygnus X massive star-forming region at $\emph{l}$ = 75.5$\degr$--87.5$\degr$. A further 37 deg$^{2}$ were observed towards targeted regions in the outer Galaxy, bringing the total survey area to 170 deg$^{2}$.

A custom source extraction algorithm, Bolocat, was designed and utilised to extract 8358 sources, with a catalogue 98 per cent complete from 0.4 to 60 Jy over all sources with object size $\leq$ 3.5$\arcmin$. The completeness limit of the survey varies as a function of longitude, with the flux density completeness limit taken as five times the median rms noise level in 1$\degr$ bins \citep{Rosolowsky10}. They concluded that the extracted sources were best described as molecular clumps-- large, dense, bound regions within which stellar clusters and large systems form.

The flux densities for each source also require a multiplication by a factor of 1.5 to provide consistency with other data sets from MAMBO and SIMBA surveys \citep{Aguirre11}.

\subsection{$^{13}$CO $J$ = 3 $\rightarrow$ 2 data}

The higher energy transition of $J$ = 3 $\rightarrow$ 2 traces higher density gas than the $J$ = 1 $\rightarrow$ 0 transition. It has a critical density of $\gtrsim$ 10$^{4}$cm$^{-3}$, compared to 10$^{2}$ -- 10$^{3}$ cm$^{-3}$ for $J$ = 1 $\rightarrow$ 0, and $\emph{E(J = 3)/k}$ = 32.8 K so is also biased towards warmer gas. $J$ = 3 $\rightarrow$ 2 is therefore less ambiguous than $J$ = 1 $\rightarrow$ 0 in identifying the emission from dense, star-forming clumps, and is useful in separating multiple emission components within a spectrum along a particular line of sight.

The $\emph{l}$ = 40$^{\circ}$ region was mapped in $^{13}$CO $J$ = 3 $\rightarrow$ 2 (330.450 GHz) with the Heterodyne Array Receiver Programme (HARP) detector at the James Clerk Maxwell Telescope (JCMT) on Mauna Kea, Hawaii. HARP has 16 receptors, each with a beam size of $\sim$ 14$\arcsec$, separated by 30$\arcsec$ and operates in the 325--375 GHz band \citep{Buckle09}. Observations were made in two parts, $\emph{l}$ = 37.83$\degr$--40.5$\degr$ in 2010 and $\emph{l}$ = 40.5$\degr$--42.5$\degr$ in 2011. The Galactic latitude range of these observations is $\mid$\emph{b}$\mid$ $\leq$ 0.5$\degr$, with a velocity range of --50 to 150 km s$^{-1}$. The increased velocity range allows for sources outside of the Solar Circle to be identified. The data were used only to provide the velocity of the peaks in the spectra extracted. The observations and reduction procedure will be discussed in more detail in a later paper.

\subsection{The VLA Galactic Plane Survey}

The VLA Galactic Plane Survey (VGPS) \citep{Stil06} mapped H$\,\textsc{I}$ and 21-cm continuum emission in Galactic longitude from $\emph{l}$ = 18$^{\circ}$ to 67$^{\circ}$ and in Galactic latitude from $\mid$\emph{b}$\mid$ $\leq$ 1.3$^{\circ}$ to $\mid$\emph{b}$\mid$ $\leq$ 2.3$^{\circ}$. The survey has an angular resolution of 1$^{\prime}$, with a spectral resolution of 1.56 km s$^{-1}$ and an rms sensitivity of 2 K. These data will be used for distance determinations, as described in Section 3.

\section{BGPS Source Distance Determination}

The $\emph{l}$ = 40$^{\circ}$ region contains 67 GRS catalogue clouds \citep{Rathborne09} in the longitude range $\emph{l}$ = 37.83$\degr$-42.50$\degr$ and latitudes $\mid$\emph{b}$\mid$ $\leq$ 0.5$\degr$ (59 with distances; \citealt{Roman-Duval09}). The upper longitude limit is set by the current extent of the HARP survey data, while the latitude range approximately corresponds to the BGPS at these longitudes. 229 BGPS sources were identified within the target area. We assigned velocities to these by extracting spectra from the HARP data cubes at the BGPS catalogue position. For sources whose spectra displayed more than one significant emission peak, the strongest emission feature was chosen \citep{Urquhart07}, which was the case for less than five per cent of sources. The BGPS actually detected sources outside the nominal latitude range, to approximately $\mid$\emph{b}$\mid$ $\leq$ 0.55$\degr$.  Five of these are included in this study. The GRS data were used as the primary indicator for those sources which fell outside of the HARP latitude range.

The positions and assigned velocities of the BGPS sources were matched to those derived for the clouds in the \citet{Rathborne09} catalogue, with positional tolerances of 5 $\times$ 5 resolution elements in the $\emph{l}$ $\times$ $\emph{b}$ directions, which corresponds to 110$\arcsec$ in each direction in the GRS data on which the \citet{Rathborne09} catalogue is based. The velocity tolerance was taken to be the full width at half-maximum of the $^{13}$CO $J$ = 1 $\rightarrow$ 0 emission line from the aforementioned catalogue. This resulted in cloud associations for 186 BGPS sources. 23 of these were associated with distance-less clouds, the remaining 163 BGPS sources were assigned the catalogued distances \citep{Roman-Duval09} of the associated GRS clouds.

This left 229 - 186 = 43 BGPS detections unassociated with GRS clouds, since no matching clouds were found in the catalogue. 9 of these sources were found to have velocities outside of the range of the GRS, by using the extended velocity range of the HARP data. These sources were assumed to lie in clouds outside the Solar Circle, probably in the Norma--Outer Arm and as such they are not catalogued by \citet{Rathborne09}. Hence no mass and no CFE is calculated for these BGPS sources. The remaining 34 sources without associated clouds but within the Solar Circle had two possible reasons for their lack of assignment: incorrect kinematic velocities, or because the host clouds were not identified by the GRS survey. Velocity assignments made initially using the $J$ = 3 $\rightarrow$ 2 data were checked against the $J$ = 1 $\rightarrow$ 0 data and found to be consistent in all cases. However, where cloud associations had not been found, we checked the velocities of any secondary (i.e.\ fainter) emission features in the $J$ = 1 $\rightarrow$ 0 spectra that were not present in the $J$ = 3 $\rightarrow$ 2 data and found a further 10 associations, with 1 of these in a distance-less cloud.

In order to look for velocity information for the remaining 24 unassociated sources, we produced velocity-integrated maps of the $^{13}$CO $J$ = 1 $\rightarrow$ 0 emission from the public GRS data. Emission was found for all 24 sources, arising in either relatively bright, compact regions, some with very small velocity ranges or from filamentary-type clouds. These likely fell below the detection criteria used by \citet{Rathborne09} of $\Delta\emph{l}$  or $\Delta\emph{b}$ $\geq$ 6$\arcmin$ or $\Delta$V $\geq$ 0.6 km s$^{-1}$. We found that 10 BGPS sources are coincident with 10 small, low velocity-dispersion clouds, with 14 BGPS sources falling in 11 filamentary-type clouds.

Table~\ref{assocs} displays a summary of the GRS cloud-BGPS associations, displaying the associations for individual clouds (only a small portion of the data is provided here, with the full list of 196 BGPS sources available as Supporting Information to the online article).

There are associated errors with these distance determinations. A full discussion of the distance determinations involved with the GRS clouds and velocity assignments can be found in \citet{Roman-Duval09}, but for clouds at a distance further than 3 kpc, the error on the kinematic distances is at most 30 per cent for the near distance and less than 20 per cent at the far distance. These uncertainties are assuming that the distance ambiguity has been correctly resolved.

\begin{table}
\begin{center}
\caption{Summary of GRS cloud parameters and BGPS source associations. Only a small portion of the data is provided here. The full list of 196 BGPS sources is available as Supporting Information to the online article.}
\label{assocs}
\begin{tabular}{lccccc} \hline
GRS Cloud &  GRS & GRS & BGPS & BGPS & BGPS \\
Name & $\emph{V$_{LSR}$}$ & $\emph{D}$ & Source & $\emph{l}$ & $\emph{b}$ \\
& (km s$^{-1}$) & (kpc) & ID & ($\degr$) & ($\degr$) \\
\hline
G039.29-00.61	&	64.55	&	4.43	&	5973	&	39.27	&	-0.59	\\
G039.29-00.61	&	64.55	&	4.43	&	5989	&	39.54	&	-0.37	\\
G039.34-00.26	&	69.65	&	4.93	&	5968	&	39.16	&	-0.17	\\
G039.34-00.26	&	69.65	&	4.93	&	5976	&	39.29	&	-0.20	\\
G039.34-00.26	&	69.65	&	4.93	&	5980	&	39.39	&	-0.14	\\
G039.34-00.26	&	69.65	&	4.93	&	5982	&	39.48	&	-0.29	\\
G039.34-00.31	&	65.82	&	4.55	&	5978	&	39.33	&	-0.32	\\
G039.34-00.31	&	65.82	&	4.55	&	5981	&	39.44	&	-0.19	\\
G039.34-00.31	&	65.82	&	4.55	&	5993	&	39.59	&	-0.21	\\
G039.49-00.21	&	17.40	&	---	&	5979	&	39.37	&	-0.18	\\
G039.49-00.21	&	17.40	&	---	&	5984	&	39.49	&	-0.18	\\
G039.49-00.21	&	17.40	&	---	&	5986	&	39.50	&	-0.20	\\
G039.49-00.21	&	17.40	&	---	&	5997	&	39.67	&	-0.16	\\
G039.59-00.01	&	43.29	&	2.85	&	5990	&	39.56	&	-0.03	\\
G039.59-00.01	&	43.29	&	2.85	&	5991	&	39.57	&	-0.04	\\
G039.59-00.01	&	43.29	&	2.85	&	5992	&	39.57	&	0.01	\\
\hline
\end{tabular}
\end{center}
\end{table}
 
The method used to assign kinematic distances to BGPS sources not associated with GRS catalogued clouds is as outlined in \citet{Eden12}. The rotation curve of \citet{Brand93} is used to assign two kinematic distances to each BGPS source. A single kinematic distance is then decided upon via the HISA method \citep[e.g.][]{Anderson09, Roman-Duval09} using H$\, \textsc{I}$ spectra from the VGPS.

Of the 24 BGPS sources not associated with GRS  catalogued clouds but found to be coincident with small clouds or filaments in the GRS data, 13 were assigned the near kinematic distance, with 11 found to be at the far kinematic distance. Table~\ref{unassoc} displays the unassociated sources and their derived kinematic distances. The clouds with which these sources are associated have no calculated CFE, since there is no CO-derived cloud mass. The CO detections of these sources have only been used to obtain LSR velocities. The current calculations of CFE contain all the GRS catalogued clouds, even those without any associated BGPS sources. Calculating a CFE that includes the small or filamentary clouds would involve producing a full catalogue of them, as well as a solid definition of what constitutes a small cloud as opposed to just an over density within the wispy CO background material.

\begin{table}
\begin{center}
\caption{The unassociated BGPS sources and their derived kinematic distances.}
\label{unassoc}
\begin{tabular}{lcccc} \hline
BGPS Source & $\emph{l}$ & $\emph{b}$ & $\emph{V$_{LSR}$}$ & $\emph{D}$\\
ID & ($\degr$) & ($\degr$) & (km s$^{-1}$) & (kpc)\\
\hline
5916	&	38.47	&	-0.07	&	24.80	&	11.62	\\
5927	&	38.67	&	0.23	&	29.22	&	11.29	\\
5955	&	38.91	&	-0.15	&	58.36	&	3.92	\\
5963	&	38.98	&	-0.08	&	16.89	&	1.13	\\
5966	&	39.05	&	0.22	&	7.11	&	12.80	\\
5969	&	39.18	&	-0.24	&	58.76	&	9.22	\\
5988	&	39.53	&	-0.20	&	51.53	&	3.47	\\
6000	&	39.69	&	-0.16	&	51.32	&	3.46	\\
6007	&	39.90	&	-0.08	&	73.43	&	7.71	\\
6010	&	39.92	&	-0.37	&	59.40	&	8.99	\\
6013	&	39.96	&	-0.15	&	57.49	&	9.12	\\
6015	&	40.07	&	0.18	&	9.29	&	12.45	\\
6020	&	40.22	&	-0.03	&	9.81	&	12.38	\\
6028	&	40.60	&	-0.10	&	65.11	&	4.58	\\
6031	&	40.74	&	0.16	&	16.29	&	1.08	\\
6033	&	40.81	&	-0.42	&	79.05	&	6.43	\\
6048	&	41.15	&	-0.08	&	49.03	&	3.35	\\
6071	&	41.51	&	-0.11	&	63.48	&	4.53	\\
6080	&	41.73	&	-0.25	&	70.28	&	5.38	\\
6081	&	41.73	&	-0.24	&	70.59	&	5.44	\\
6083	&	41.76	&	-0.06	&	26.50	&	1.80	\\
6086	&	41.88	&	0.47	&	20.55	&	11.28	\\
6087	&	41.88	&	0.49	&	21.19	&	11.23	\\
6096	&	42.10	&	0.35	&	20.76	&	1.39	\\
\hline
\end{tabular}
\end{center}
\end{table}

The molecular clouds associated with 196 BGPS sources have been identified. In Table~\ref{total} we present the GRS clouds with the number of associated BGPS sources. There are 67 molecular clouds from the \citet{Rathborne09} catalogue in the $\emph{l}$ = 40$^{\circ}$ region. We have associated 47 of these clouds with 196 BGPS sources, with only nine having just a single associated BGPS source.

\begin{table*}
\begin{center}
\caption{Summary of GRS cloud parameters, number of BGPS source associations and the associated BGPS source masses.}
\label{total}
\begin{tabular}{lcccccccccc} \hline
 GRS Cloud & $\emph{l}$ & $\emph{b}$ & $\emph{V$_{LSR}$}$ & $\emph{M$_{cloud}$}$ & $\emph{D}$ & No. BGPS & $\emph{$M_{clumps}$}$ & CFE & $\Delta$CFE & Arm/\\
Name & ($\degr$) & ($\degr$) & (km s$^{-1}$) & (M$_{\odot}$) & (kpc) & Sources & (M$_{\odot}$) & (\%) & (\%) & Inter-arm\\
 \hline
G037.59-00.66	&	37.59	&	-0.66	&	20.76	&	2150	&	1.45	&	0	&	0	&	0.0	&	0.0	&	a	\\
G037.69+00.09	&	37.69	&	0.09	&	84.10	&	297000	&	6.70	&	5	&	5830	&	2.0	&	1.0	&	i	\\
G037.74-00.06	&	37.74	&	-0.06	&	86.65	&	290000	&	6.70	&	11	&	7288	&	2.5	&	0.4	&	i	\\
G037.74-00.46	&	37.74	&	-0.46	&	74.75	&	17800	&	5.25	&	0	&	0	&	0.0	&	0.0	&	a	\\
G037.74+00.19	&	37.74	&	0.19	&	45.40	&	---	&	---	&	0	&	---	&	---	&	---	&	---	\\
G037.79+00.24	&	37.79	&	0.24	&	47.54	&	9160	&	10.32	&	0	&	0	&	0.0	&	0.0	&	i	\\
G037.84-00.41	&	37.84	&	-0.41	&	64.97	&	106000	&	9.05	&	1	&	554	&	0.5	&	0.3	&	a	\\
G037.89-00.21	&	37.89	&	-0.21	&	13.54	&	1400	&	1.05	&	0	&	0	&	0.0	&	0.0	&	a	\\
G037.89-00.41	&	37.89	&	-0.41	&	61.15	&	134000	&	9.32	&	2	&	15000	&	11.2	&	4.6	&	a	\\
G038.04-00.26	&	38.04	&	-0.26	&	13.11	&	145	&	1.02	&	0	&	0	&	0.0	&	0.0	&	a	\\
G038.04+00.19	&	38.04	&	0.19	&	42.87	&	382	&	2.80	&	1	&	371	&	97.1	&	52.6	&	i	\\
G038.19-00.16	&	38.19	&	-0.16	&	62.85	&	34900	&	4.22	&	3	&	876	&	2.5	&	1.2	&	a	\\
G038.24-00.16	&	38.24	&	-0.16	&	65.40	&	125000	&	8.93	&	5	&	4920	&	3.9	&	1.6	&	a	\\
G038.49+00.14	&	38.49	&	0.14	&	17.36	&	1380	&	1.27	&	8	&	133	&	9.6	&	4.5	&	a	\\
G038.54-00.06	&	38.54	&	-0.06	&	16.94	&	1370	&	1.25	&	7	&	87	&	6.4	&	2.9	&	a	\\
G038.59-00.41	&	38.59	&	-0.41	&	19.06	&	1210	&	1.35	&	3	&	67.5	&	5.6	&	2.7	&	a	\\
G038.69-00.06	&	38.69	&	-0.06	&	36.07	&	1700	&	2.38	&	2	&	49.6	&	2.9	&	1.6	&	i	\\
G038.69-00.11	&	38.69	&	-0.11	&	42.02	&	619	&	2.75	&	0	&	0	&	0.0	&	0.0	&	i	\\
G038.69+00.44	&	38.69	&	0.44	&	16.94	&	215	&	1.25	&	0	&	0	&	0.0	&	0.0	&	a	\\
G038.74-00.46	&	38.74	&	-0.46	&	50.52	&	20800	&	9.93	&	5	&	9220	&	44.3	&	17.6	&	i	\\
G038.79-00.51	&	38.79	&	-0.51	&	66.25	&	128000	&	8.70	&	7	&	6310	&	4.9	&	2.0	&	a	\\
G038.89-00.26	&	38.89	&	-0.26	&	44.14	&	1850	&	2.90	&	3	&	374	&	20.2	&	7.2	&	i	\\
G038.94-00.46	&	38.94	&	-0.46	&	41.59	&	488000	&	10.50	&	15	&	53000	&	10.9	&	2.7	&	i	\\
G038.99-00.41	&	38.99	&	-0.41	&	60.72	&	297000	&	9.12	&	3	&	2290	&	0.8	&	0.3	&	a	\\
G039.09+00.49	&	39.09	&	0.49	&	22.89	&	32000	&	11.60	&	2	&	3750	&	11.7	&	5.7	&	a	\\
G039.19+00.49	&	39.19	&	0.49	&	28.80	&	---	&	---	&	0	&	---	&	---	&	---	&	---	\\
G039.24-00.61	&	39.24	&	-0.61	&	16.50	&	---	&	---	&	3	&	---	&	---	&	---	&	---	\\
G039.24-00.06	&	39.24	&	-0.06	&	22.46	&	22000	&	11.60	&	2	&	28310	&	128.7	&	55.0	&	a	\\
G039.29-00.61	&	39.29	&	-0.61	&	64.55	&	16100	&	4.43	&	2	&	664	&	4.1	&	2.1	&	a	\\
G039.34-00.26	&	39.34	&	-0.26	&	69.65	&	34300	&	4.93	&	4	&	1770	&	5.2	&	2.0	&	a	\\
G039.34-00.31	&	39.34	&	-0.31	&	65.82	&	25400	&	4.55	&	3	&	1290	&	5.1	&	2.2	&	a	\\
G039.49+00.29	&	39.49	&	0.29	&	42.44	&	301	&	2.80	&	0	&	0	&	0.0	&	0.0	&	i	\\
G039.49-00.21	&	39.49	&	-0.21	&	17.40	&	---	&	---	&	4	&	---	&	---	&	---	&	---	\\
G039.54+00.29	&	39.54	&	0.29	&	15.20	&	---	&	---	&	0	&	---	&	---	&	---	&	---	\\
G039.59-00.01	&	39.59	&	-0.01	&	43.29	&	778	&	2.85	&	3	&	117	&	15.0	&	8.7	&	i	\\
G039.69-00.56	&	39.69	&	-0.56	&	83.25	&	49500	&	6.55	&	2	&	1010	&	2.0	&	1.0	&	i	\\
G039.89-00.21	&	39.89	&	-0.21	&	57.80	&	---	&	---	&	14	&	---	&	---	&	---	&	---	\\
G040.09-00.51	&	40.09	&	-0.51	&	57.75	&	329000	&	9.10	&	4	&	3270	&	1.0	&	0.3	&	a	\\
G040.29+00.19	&	40.29	&	0.19	&	82.82	&	5040	&	6.47	&	1	&	266	&	5.3	&	3.6	&	i	\\
G040.34-00.26	&	40.34	&	-0.26	&	72.20	&	51100	&	5.43	&	5	&	3950	&	7.7	&	2.6	&	a	\\
G040.84-00.16	&	40.84	&	-0.16	&	23.74	&	757	&	1.65	&	5	&	160	&	21.1	&	10.1	&	a	\\
G040.89-00.21	&	40.89	&	-0.21	&	26.70	&	---	&	---	&	1	&	---	&	---	&	---	&	---	\\
G040.99+00.04	&	40.99	&	0.04	&	74.75	&	14800	&	6.40	&	0	&	0	&	0.0	&	0.0	&	i	\\
G041.04-00.26	&	41.04	&	-0.26	&	39.04	&	435000	&	10.23	&	6	&	4220	&	1.0	&	0.3	&	i	\\
G041.04-00.26	&	41.04	&	-0.26	&	65.82	&	15800	&	4.72	&	4	&	1070	&	6.8	&	2.3	&	a	\\
G041.04-00.51	&	41.04	&	-0.51	&	75.17	&	2430	&	6.38	&	0	&	0	&	0.0	&	0.0	&	i	\\
G041.19-00.21	&	41.19	&	-0.21	&	59.87	&	291000	&	8.65	&	14	&	22600	&	7.8	&	1.6	&	a	\\
G041.24-00.56	&	41.24	&	-0.56	&	75.60	&	6930	&	6.40	&	0	&	0	&	0.0	&	0.0	&	i	\\
G041.24+00.39	&	41.24	&	0.39	&	71.35	&	2830	&	5.53	&	1	&	118	&	4.2	&	3.0	&	a	\\
G041.29+00.34	&	41.29	&	0.34	&	14.81	&	7280	&	11.65	&	2	&	3360	&	46.2	&	22.1	&	a	\\
G041.34-00.16	&	41.34	&	-0.16	&	13.54	&	10700	&	11.70	&	4	&	4370	&	40.8	&	15.5	&	a	\\
G041.34+00.09	&	41.34	&	0.09	&	60.30	&	67100	&	8.55	&	6	&	4600	&	6.9	&	2.2	&	a	\\
G041.59+00.29	&	41.59	&	0.29	&	59.02	&	2240	&	4.10	&	0	&	0	&	0.0	&	0.0	&	a	\\
G041.74+00.04	&	41.74	&	0.04	&	17.79	&	56600	&	11.38	&	4	&	2950	&	5.2	&	2.6	&	a	\\
G041.79+00.49	&	41.79	&	0.49	&	41.17	&	71500	&	9.93	&	0	&	0	&	0.0	&	0.0	&	i	\\
G042.04-00.01	&	42.04	&	-0.01	&	57.75	&	243000	&	8.60	&	8	&	9070	&	3.7	&	1.0	&	a	\\
G042.04+00.19	&	42.04	&	0.19	&	18.21	&	3740	&	11.30	&	0	&	0	&	0.0	&	0.0	&	a	\\
G042.09-00.11	&	42.09	&	-0.11	&	16.51	&	201	&	1.23	&	3	&	50.4	&	25.1	&	12.4	&	a	\\
G042.14-00.61	&	42.14	&	-0.61	&	67.52	&	286000	&	5.12	&	1	&	251	&	0.1	&	0.0	&	a	\\
G042.14+00.09	&	42.14	&	0.09	&	15.66	&	83.8	&	1.17	&	1	&	45.9	&	54.8	&	28.1	&	a	\\
G042.19-00.61	&	42.19	&	-0.61	&	34.79	&	1880	&	2.33	&	1	&	110	&	5.9	&	2.5	&	i	\\
G042.29-00.51	&	42.29	&	-0.51	&	75.60	&	18400	&	6.28	&	1	&	465	&	2.5	&	1.5	&	i	\\
G042.34+00.39	&	42.34	&	0.39	&	15.24	&	6050	&	11.40	&	0	&	0	&	0.0	&	0.0	&	a	\\
G042.34-00.31	&	42.34	&	-0.31	&	27.60	&	---	&	---	&	2	&	---	&	---	&	---	&	---	\\
G042.44-00.46	&	42.44	&	-0.46	&	10.99	&	754	&	0.93	&	0	&	0	&	0.0	&	0.0	&	a	\\
G042.44-00.26	&	42.44	&	-0.26	&	65.40	&	140000	&	4.90	&	2	&	2100	&	1.5	&	0.6	&	a	\\
G042.44-00.51	&	42.44	&	-0.51	&	33.94	&	898	&	2.28	&	0	&	0	&	0.0	&	0.0	&	i	\\
\hline
\end{tabular}
\end{center}
\end{table*}

\begin{figure}
\begin{center}
\includegraphics[scale=0.5]{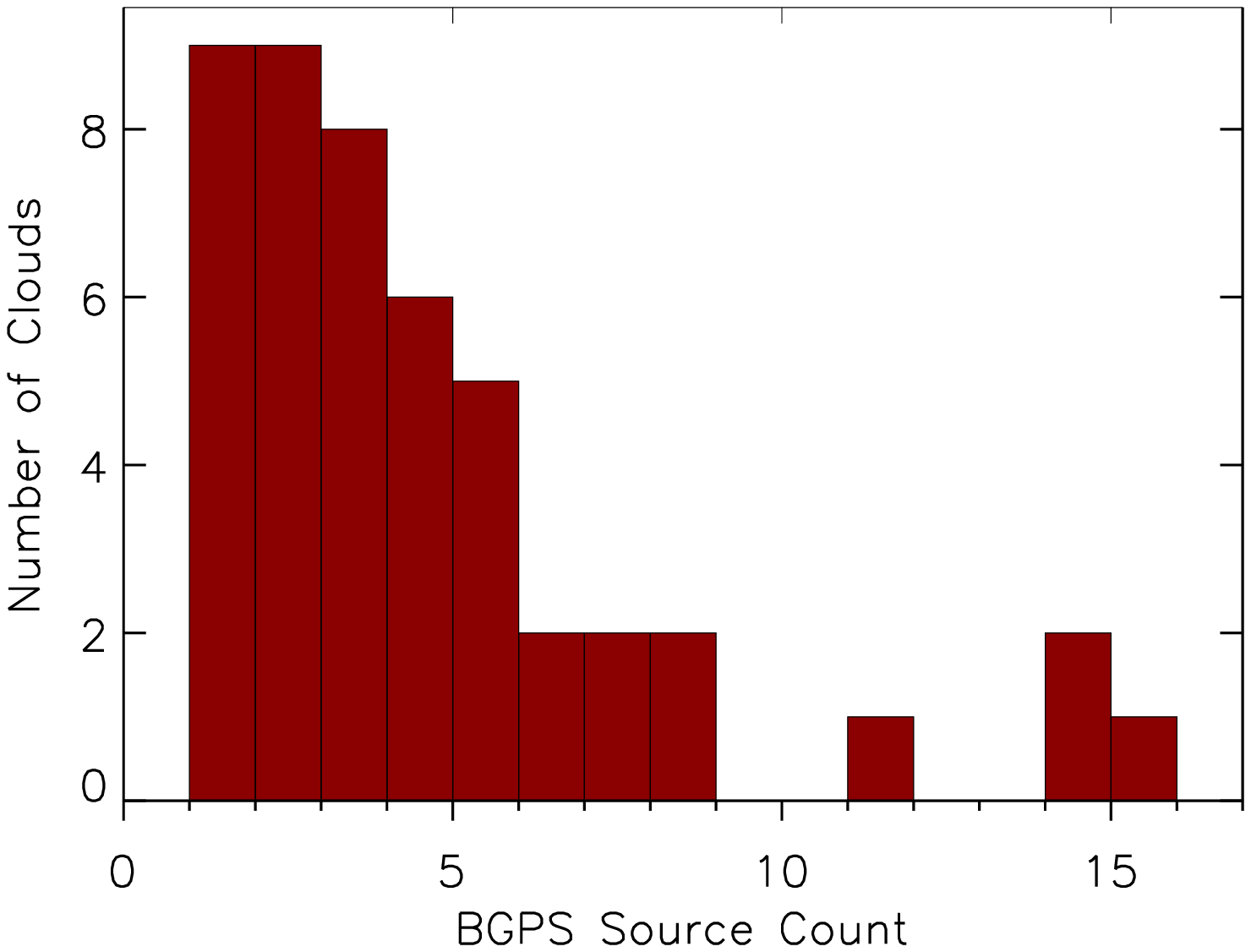}
\caption{The number of BGPS sources found in each molecular cloud with at least one association.}
\label{distri}
\end{center}
\end{figure}

In Fig.~\ref{distri} we present a histogram of the number of BGPS sources found in each GRS molecular cloud. All but 4 of these clouds are associated with less than 10 BGPS sources. Of the clouds with 10 or more, 2 are found in the inter-arm regions, 1 is in the Sagittarius arm and the fourth has no distance determination.

\section{Results and Analysis}

\subsection{Determining inter-arm clouds}

In aiming to test the effect of Galactic structure on the star-formation process, it is first important to distinguish between the spiral arms and inter-arm regions. The separation of these components is not an exact process, but the main aim is to ensure that there is sufficient separation such that the emission in each of the structure bins is dominated by sources within the main structures.

Models of the spiral structure of the Milky Way show evidence for both a two-armed model \citep{Francis12} and a four-armed model \citep{Vallee95, Hou09}. However, the model of \citet{Vallee95} is chosen to describe the spiral arm geometry in this study as this is the model used by \citet{Roman-Duval09} in determining the Galactic distribution of the GRS molecular gas, which points towards a four-armed model. \citet{Hou09} also find evidence that a four-armed model is favoured due to the distribution of H$\,\textsc{II}$ regions and GMCs.

By using the model of \citet{Vallee95}, the loci of the spiral arms (Scutum--Centaurus, Sagittarius and Perseus) are described by the following equations:

\begin{equation}
r = 2.65 \; \rmn{kpc} \; \rmn{e}^{(\theta + \theta_{0})\rmn{tan}(p)}
\end{equation}

\noindent where $r$ is the Galactocentric radius, $\theta$ is the azimuth around the Galactic Centre with origin located on the Galactic Centre-Sun axis, $\theta_{0}$ = $\pi$, $3\pi/2$ and $2\pi$ for the Scutum--Centaurus, Sagittarius and Perseus arms, respectively, and $p$ is the pitch angle and is equal to 12.7$\degr$.

\begin{figure}
\includegraphics[scale=0.5]{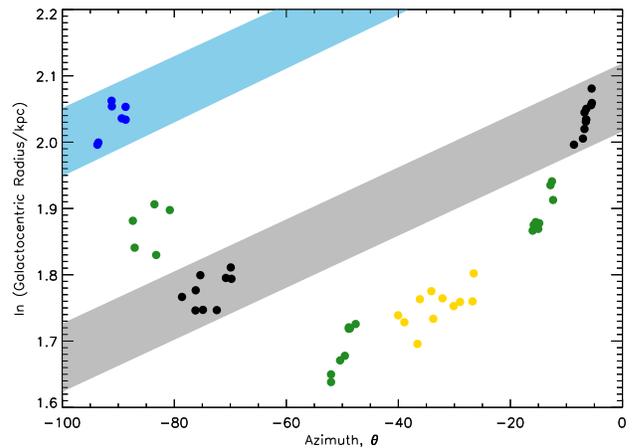}
\caption{Position of the GRS clouds in ($\theta$, ln($r$)) space, where the spiral arms are represented by straight lines. The positions of the arms are determined by the model of \citet{Vallee95}, with the blue and grey bands representing the taken extent of the Perseus, and Sagittarius arms, respectively. The clouds which fall in those arms are marked with blue and black circles, respectively, whilst the inter-arm clouds are represented with the green circles and the yellow circles are the Scutum--Centaurus tangent clouds.}
\label{azimuth}
\end{figure}

Making use of the kinematic velocities of each cloud in the region \citep{Rathborne09} and the distances to the GRS clouds \citep{Roman-Duval09}, the galactocentric radius and azimuth of each cloud can be calculated. Fig.~\ref{azimuth} displays the position of each cloud in ($\theta$, ln($r$)) space, where the spiral arms are represented by straight lines. The populations that correspond to the spiral arm components are coloured the same as the lines, with the inter-arm components represented by the green circles. The clouds at $\sim$ -50$\degr$ azimuth are those at the $\emph{l}$ = 40$\degr$ tangent point, which is the Sagittarius--Scutum--Centaurus inter-arm zone (see also \citealt{Sawada12}). The Scutum--Centaurus tangent clouds (those indicated by the yellow circles) do not fall on the Scutum--Centaurus arm, as indicated by the model, but are considered to be in the arm as they have the velocity distribution and distance that corresponds to the Scutum--Centaurus tangent in this line of sight. The models of the arms used here do not take into account the confusion that occurs in the bar-end/Scutum--Centaurus tangent region due to the streaming motions and large range of velocities found in this environment. Clouds were placed in the Perseus and Sagittarius arms if they fell within 0.5 kpc of the taken position of the spiral arm from Equation 1.

\subsection[]{Masses of BGPS sources}

\begin{figure}
\begin{center}
\includegraphics[scale=0.5]{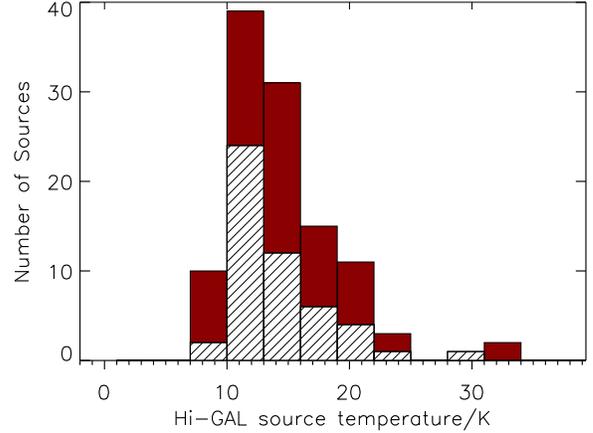}
\caption{The distributions of SED-based temperatures of the matched Hi-GAL--BGPS sources for the $\emph{l}$ = 40$\degr$ region with the arm and spiral arm components depicted by the red and white hashed bars respectively. The regions were separated using the distances derived to the matched BGPS sources.}
\label{Hi-GAL}
\end{center}
\end{figure}

The 1.1-mm dust continuum flux densities from BGPS are converted to source masses using the standard formula:

\begin{equation}
M = \frac{S_{\nu}D^{2}}{\kappa_{\nu}B_{\nu}(T_{d})}
\end{equation}

\noindent which leads to:

\begin{equation}
M=13.1 M_{\odot} \left(\frac{S_{\nu}}{1 Jy}\right)\left(\frac{D}{1 kpc}\right)^{2}(\rmn{e}^{13.12/T_{d}}-1)
\end{equation}

\noindent where $\kappa_{\nu}$ = 0.0114 cm$^{2}$ g$^{-1}$, $S_{\nu}$ is the catalogued flux density, $D$ is the source distance and $B_{\nu}$ is the Planck function evaluated at dust temperature $T_{d}$. A single temperature of 14 K is used for all the sources in the $\emph{l}$ = 40$\degr$ region, in contrast to the value of 20 K assumed by \citet{Dunham10}. The lower value is supported by the distribution of source temperatures in the $\emph{l}$ = 40$\degr$ region found from 70 -- 500 $\mu$m photometry extracted from the Herschel infrared Galactic-plane survey (Hi-GAL; \citealt{Molinari10}) and confirmed by \citet{Veneziani12} who found lower temperatures in the $\emph{l}$ = 59$\degr$ field, a field similar to the $\emph{l}$ = 40$\degr$ region. The BGPS sources were cross-matched with Hi-GAL detections (Schisano et al, in preparation), and the temperatures derived by spectral energy distribution (SED) and greybody fitting (D. Elia, private communication). Fig.~\ref{Hi-GAL} shows that the peak of the temperature distribution lies at $\sim$ 14 K in both the arm and inter-arm components of the $\emph{l}$ = 40$\degr$ region and this represents the most probable temperature.  Allowing for a 1-$\sigma$ spread of the peak, clump masses may be over-estimated by a factor of 1.7 or under-estimated by 0.7. A Kolomogorov--Smirnov (K--S) test was applied to the $\emph{T}_{d}$ distributions of the two sub-samples. This K--S test showed that they could be assumed to have the same temperatures as there was an 88 per cent probability that the temperatures were from the same sample. Observations of M83 by \citet{Foyle12} found higher dust temperatures in the spiral arms compared to the inter-arm regions, in contrast to our results. This difference could be due to tracing a different dust mass component. At low resolutions, \citet{Foyle12} may include a large, low-density component which is heated by the interstellar radiation field, whereas the BGPS and Hi-GAL trace dense cores that are mostly shielded by high extinction. \citet{Paradis12} calculated the dust temperature and 500 $\mu$m emissivity excess as a function of Galactic longitude, with peaks correlating with the locations of Galactic spiral arms. However, this is at resolutions of 4$\arcmin$ and will include diffuse emission, while we are using extracted point sources and excluding the diffuse material.

It should also be noted that absolute masses and accurate distances are not vital to most of the results in this study since we are mainly concerned with mass ratios (Section 4.3) in which case uncertainties in distance are not an issue. Reliable associations between BGPS sources and molecular clouds are much more important and the separation into the different populations identified with Galactic structures can be achieved with source velocities alone.

\subsection{Clump formation efficiencies}

The clump formation efficiency (CFE) is a measure of the fraction of molecular gas that has been converted into dense, potentially star-forming clumps. This quantity is analogous (or is a precursor) to the star formation efficiency. The CFE must be viewed as an upper limit to the SFE or the first step in a sequence of conversion efficiencies from molecular clouds to stars.

The CFE is a measure of: 

\begin{equation}
\frac{M_{clump}}{M_{cloud}}= \frac{1}{M_{cloud}} \int^t_0 \frac{dM}{dt}\,dt
\end{equation}

\noindent where $\emph{dM/dt}$ is the instantaneous clump formation rate. A high value for the CFE can indicate either a high clump formation rate or a long formation timescale.

Using the catalogued GRS cloud masses \citep{Roman-Duval10} and the derived masses for the BGPS sources, we are able to calculate total CFEs for the different velocity components of the $\emph{l}$ = 40$\degr$ field. The total CFEs also include the masses of the unassociated BGPS sources, which were assigned to populations by the same method used to assign clouds as they have velocities and derived kinematic distances, and the GRS clouds with no associated BGPS sources. The total CFE values for the spiral arm and inter-arm regions respectively are 5.5 $\pm$ 0.6 and 4.9 $\pm$ 0.7 per cent. The individual spiral arms had total CFEs of 2.0 $\pm$ 0.4, 4.3 $\pm$ 0.5 and 36.3 $\pm$ 9.0 per cent for the Scutum--Centaurus, Sagittarius and Perseus arms, respectively, with the individual inter-arm regions having CFEs of 15.8 $\pm$ 2.9, 2.3 $\pm$ 0.5 and 6.5 $\pm$ 1.0 per cent. These inter-arm regions are listed in the order of decreasing azimuthal angle, as indicated in Fig.~\ref{azimuth}. These values do not include the masses of the small, low velocity-dispersion clouds discussed in Section 3.1 that were not included by \citet{Rathborne09}, and as such the CFEs can be taken as an upper limit. The uncertainties on the CFEs are a combination of the catalogued GRS cloud mass uncertainties \citep{Roman-Duval10}, the uncertainties in BGPS flux densities \citep{Rosolowsky10} and the distribution of source temperatures, using the variance of the distribution. Any biases corresponding to the distance distribution of cloud masses are discussed in detail by \citet{Roman-Duval10}. The total CFE values obtained for spiral-arm and inter-arm clouds are consistent within the uncertainties. In the separated spiral-arm components, the CFE in the Perseus arm shows a large increase, significant at the 3-$\sigma$ level.

The CFE calculations are based on the molecular mass accounted for in the catalogue of \citet{Roman-Duval10} and all the BGPS-traced mass with a known velocity, both with and without an associated CO mass. The 24 BGPS sources without an associated cloud make up 12 per cent of the 196 sources and only 6 per cent of the total clump mass. The molecular mass not counted in the filamentary clouds and small, low velocity-dispersion clouds makes up some small part of the 37 per cent of the total molecular mass in the GRS data not picked up by the CLUMPFIND search of \citet{Roman-Duval09}. As a result, combining these two effects, we can say that inclusion of the clump masses without an associated molecular mass does not significantly bias the CFE values.

\section{Discussion}

\subsection{Inter-arm star formation}

Determining the quantity of ongoing star formation in the inter-arm regions is key to understanding how important the spiral arms are in the overall production of stars and what effect, if any, they have on the star-formation rate (SFR). Studies of other galaxies have shown higher star-formation efficiencies in the spiral arms compared to inter-arm regions in H$\alpha$ fluxes and H$\,\textsc{I}$ \citep[e.g.][]{Cepa90,Seigar02}. However, \citet{Bigiel08} and \citet{Blanc09} have argued that the SFR correlates with CO intensity and the surface density of the molecular gas, as opposed to the H$\,\textsc{I}$ component.

The total SFR in a molecular cloud or galaxy is directly related to, and possibly determined by, the amount of dense molecular gas it contains. SFRs and total molecular masses are correlated over nine orders of magnitude in mass scale and can be described by a family of linear scaling laws, parameterised by the fraction of molecular gas that is dense (i.e. $n(H_{2}) > 10^{4}$ cm$^{-3}$). That is, the underlying star-formation scaling law is always linear for systems with the same dense gas fraction \citep{Lada12}. It is also shown that the SFR is linear on scales from Galactic GMCs to sub-millimeter galaxies above a density threshold \citep{Krumholz12}.

\begin{figure}
\begin{center}
\includegraphics[scale=0.5]{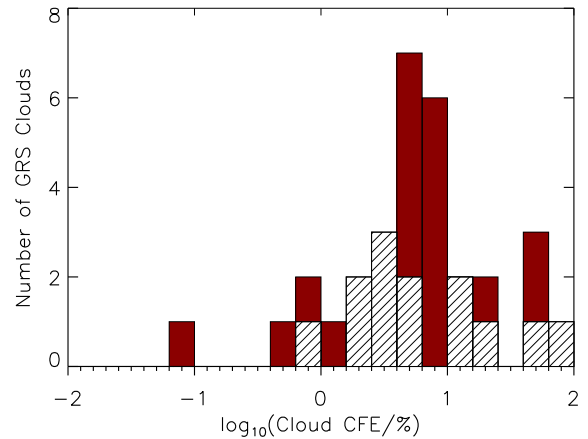}
\caption{Distribution of the clump formation efficiencies for individual GRS clouds with the spiral arm and inter-arm components depicted by the red and white hashed bars respectively.}
\label{CFEhist}
\end{center}
\end{figure}

By defining the star-formation efficiency in terms of the ratio of SFR, derived from both the far-ultraviolet and 24 $\mu$m emission, and the H$_{2}$ surface density, \citet{Foyle10} found no enhancement in the molecular fraction of total gas mass, nor in the star-formation efficiency compared to inter-arm regions, in the arms of two external spiral galaxies, with SFE variations found to be set by local environmental factors \citep{Bigiel08}. By taking the CFE to be as analogous to the SFE, we can say that this is consistent with the results of this study. The median cloud CFEs are 5.6 $\pm$ 3.1 and 5.3 $\pm$ 3.3 per cent for the arm and inter-arm regions, respectively, where the uncertainties are the median absolute deviations. The mean cloud CFE's are 14.9 $\pm$ 4.8 and 16.3 $\pm$ 7.5 per cent, respectively. The two samples are displayed in Fig.~\ref{CFEhist} and a K--S test shows that there is a 58 per cent probability that they are from the same population.  So, neither the total CFEs, or those for the individual groups of clouds show any evidence of systematic difference with Galactic environment.

However, there are significant variations from cloud to cloud, as seen in Fig.~\ref{CFEhist}, but they are part of a single population. This is consistent with \citet{Eden12} and suggests that local feedback on the scale of individual clouds is the dominant process in determining CFE or SFE changes.

The total CFEs for the inter-arm regions indicate that there is both inefficient and efficient star formation going on in the inter-arm regions as well as in the spiral arms. CFEs in the individual clouds, as shown in Fig.~\ref{CFEhist}, show that high-CFE clouds are found both in the inter-arm and spiral arms. Three of the five clouds with a CFE found to be greater than 40 per cent are associated with a H$\,\textsc{II}$ region. The presence of these H$\,\textsc{II}$ regions could be the cause of the increase due to the correlation between feedback processes and an increase in star-formation efficiency \citep[e.g.][]{Moore07}.

There is evidence that the $\emph{l}$ = 40$\degr$ line of sight is host to a Galactic spur \citep{Weaver70,Shane72} between the Sagittarius and Scutum arms. These inter-arm spurs are observed in external galaxies \citep[e.g.][]{Corder08,Muraoka09}, with the spurs in M51 and M83 well correlated with inter-arm H$\,\textsc{II}$ regions and sites of massive star formation. This region, which we have counted as inter-arm, is found to have a much lower CFE than the total inter-arm region at 2.3 $\pm$ 0.5 per cent. Thus its inclusion cannot bias the results in the sense of producing an artificially high CFE for the inter-arm zones. \citet{Sakamoto97} found that there was a lower than average gas density in this region, implying that this spur is not of a similar gas density to spiral arm gas, and no enhancement compared to other inter-arm regions.

\subsection{Star formation in the Perseus arm}

\citet{Moore12} found that the infrared luminosity-to-cloud mass ratio, in the sector of the Galactic plane covered by the GRS, was significantly increased in the Perseus arm, compared to the Scutum--Centaurus tangent, the Sagittarius arm and adjacent inter-arm regions. However, this increase could be entirely accounted for by the presence of the W49A massive star-forming region, a promising Galactic analogue for an extragalactic starburst system with dust temperatures $>$ 100 K and densities $>$ 10$^{5}$ cm$^{-3}$ \citep{Nagy12}.

\begin{figure}
\begin{center}
\includegraphics[scale=0.5]{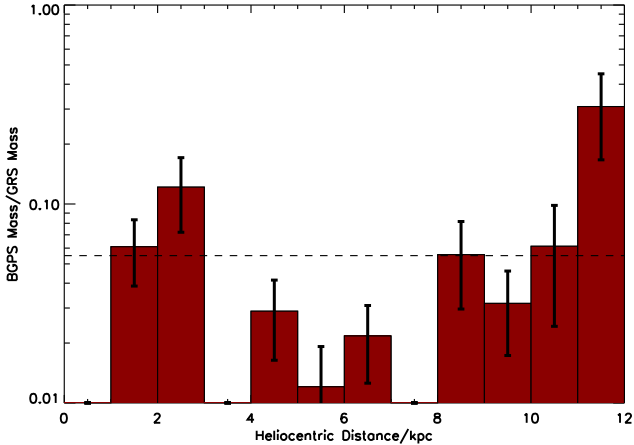}
\caption{The total CFE as a function of heliocentric distance within the $\emph{l}$ = 40$\degr$ region, with the Perseus arm found at 11.0-12.0 kpc. The dashed line indicates the total CFE for spiral arm regions.}
\label{Perseus}
\end{center}
\end{figure}

The $\emph{l}$ = 40$\degr$ region is a subset of the area studied in \citet{Moore12} but W49A is not included here. However, we do still find a peak in CFE in the Perseus arm clouds, as can be seen in Fig.~\ref{Perseus}. The total CFE for spiral arm regions is marked with the dashed line. There is a peak in the CFE at the heliocentric distance associated with the Perseus arm (11.0-12.0 kpc at $\emph{l}$ = 40$\deg$) and the total CFE for the Perseus arm clouds is found to depart at the 3--$\sigma$ level from that in the other zones, with a CFE of 36.3 $\pm$ 9.0 per cent. The uncertainties are the uncertainties described in Section 4.3, with a Poisson consideration on the number of GRS clouds. This increase is possibly still the effect of W49A ($\emph{l}$ = 43.17$\degr$ which falls outside the $\emph{l}$ = 40$\degr$ region), but this would require it to be doing so at a scale of $\sim$ 800 pc, if we assume a distance of 11.4 kpc \citep{Gwinn92}. Increases in the CFE related to triggering from feedback have been observed \citep{Moore07} but only on the scale of tens of parsecs. It is unlikely that an individual star-forming region would have an effect at this distance although it may be that the peripheral clouds around W49A itself may also have abnormally high CFE.

As mentioned in Section 4.3, the high CFE implies either a high clump-formation rate or a long formation timescale. However, as the clump stage is shown to be short, $\sim$ 0.5 Myr \citep[e.g.][]{Ginsburg12}, this increase is more likely to be due to an increased formation rate in this spiral arm as it would otherwise require much more extended dense-clump lifetimes, for which there is no evidence.

\citet{Roman-Duval10} suggest that the GRS survey may be under-sampled at the distance of the Perseus arm, mainly because the arm was not as clearly defined as other arms in the cloud distribution. If this were the case, it would result in artificially elevated CFE values and bring the result discussed above into question. However, there is no direct evidence for under-sampling and other data at similar Galactocentric radii have also revealed spiral arms that are poorly-defined in star-formation tracers (e.g. Urquhart et al. 2013, in prep). The BGPS traces mass further than the catalogued clouds of the GRS, however, by removing the sources with a heliocentric distance further than the furthest GRS cloud, the Perseus CFE still remains elevated at 31 per cent.

\subsection{The scales of star formation}

The formation of stars is an evolutionary and hierarchical process. Molecular clouds form from atomic gas and these clouds form internal dense clumps within which cluster sized systems form, and which in turn house cores where single stars or small multi-star systems are produced. Each of these stages is subject to its own efficiency, each of which can be measured. The formation efficiency of molecular clouds can be obtained from the ratio of H$_{2}$ to H$\,\textsc{I}$ mass, that of the star-forming clumps from clouds is measured by the CFE  discussed here and the conversion of stars out of the gas from the infrared luminosity-to-cloud or clump mass.

\begin{figure}
\begin{center}
\includegraphics[scale=0.5]{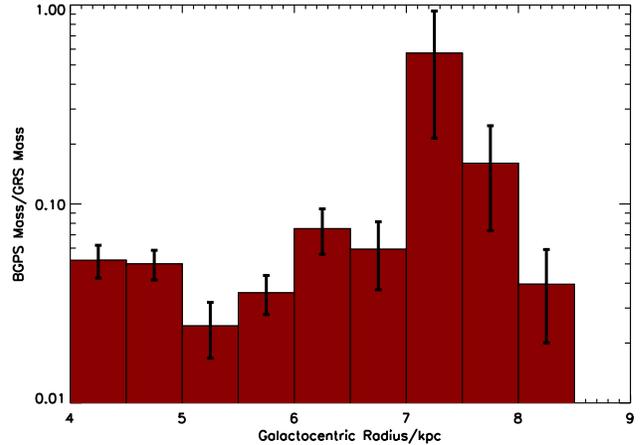}
\caption{The total CFE as a function of galactocentric radius within the $\emph{l}$ = 30$\degr$ region, outlined in \citet{Eden12}, and the $\emph{l}$ = 40$\degr$ region. Sources with heliocentric distance less than than 2 kpc are omitted, in order to remove local sources that might affect the results at galactocentric radii of $\sim$ 8 kpc.}
\label{GRCFE}
\end{center}
\end{figure}

If the environment pertaining to large-scale structure were changing the star formation process, at least one of these efficiencies should show some variation with environment on kpc scales. The work of \citet{Foyle10} has shown no variation between the inter-arm and arm regions in the ratio of molecular-to-atomic gas, albeit only in two spiral galaxies. This study, as well as that of \citet{Eden12}, has shown no evidence that the CFE is dependent on proximity to spiral arms or varies between arms. Fig.~\ref{GRCFE} displays the CFE as a function of galactocentric radius, combining the results of this study and \citet{Eden12}. However, it has been found that the ratio of the integrated YSO luminosity to molecular cloud mass is influenced by the presence of some, but not all, spiral-arm structures \citep{Moore12}. This implies that the clump-to-stars stage is affected by spiral structure, but it is unclear as to how this occurs.

\section{Conclusions}

By associating 196 BGPS sources to GRS clouds with known distances, and using the rotation curve of \citet{Brand93} combined with kinematic velocities, we assigned kinematic distances to 196 BGPS sources found in the Galactic Plane slice from $\emph{l}$ = (37.83$\degr$-42.50$\degr$), $\mid$\emph{b}$\mid$ $\leq$ 0.5$\degr$.

The distances and kinematic velocities of the GRS clouds \citep{Rathborne09,Roman-Duval09} were inverted to give galactocentric radii and azimuthal angles, allowing for clouds to be located in terms of arm and inter-arm material, and by separating the clouds into these two groups, we are able to test how the clump formation efficiency (CFE) varies with Galactic environment.

The CFEs, defined as the clump-to-cloud mass ratio, were found to be 5.5 $\pm$ 0.56 and 4.9 $\pm$ 0.7 per cent for the combined arm and inter-arm regions, respectively, hence consistent with each other. The CFEs for the individual arms 2.0 $\pm$ 0.4, 4.3 $\pm$ 0.5 and 36.3 $\pm$ 9.0 per cent for the Scutum--Centaurus, Sagittarius and Perseus arms, respectively.

The median cloud CFEs for the arm and inter-arm regions are 5.6 $\pm$ 3.1 and 5.3 $\pm$ 3.3 per cent, respectively, with corresponding mean values of 14.9 $\pm$ 4.8 and 16.3 $\pm$ 7.5 per cent. These are also consistent with each other.

The work of \citet{Foyle10}, \citet{Eden12} and \citet{Moore12}, combined with this study shows that that the large-scale structure does not seem to change the efficiency of the formation of the clouds or clumps from which stars form and any increases in SFE may come from the conversion of clumps to stars.

\section*{Acknowledgments}

This publication makes use of molecular line data from the Boston University-FCRAO Galactic Ring Survey (GRS). The GRS is a joint project of Boston University and Five College Radio Astronomy Observatory, funded by the National Science Foundation under grants AST-9800334, AST-0098562, \& AST-0100793. The James Clerk Maxwell Telescope is operated by The Joint Astronomy Centre on behalf of the Science and Technology Facilities Council of the United Kingdom, the Netherlands Organisation for Scientific Research, and the National Research Council of Canada. The National Radio Astronomy Observatory is a facility of the National Science Foundation operated under cooperative agreement by Associated Universities, Inc. DJE wishes to acknowledge an STFC PhD studentship for this work.

\bibliographystyle{mn2e}
\bibliography{ref}

\label{lastpage}

\end{document}